\documentstyle[12pt,epsfig]{article}

\newcommand{\be}{\begin{eqnarray}}
\newcommand{\ee}{\end{eqnarray}}

\def\anue{{\bar\nu_e}}

\newcommand{\dm}{\mbox{$\Delta m_{21}^2$~}}

\newcommand{\kl}{\mbox{KamLAND~}}

\newcommand{\sss}{\sin^2 \theta_{12}}
\newcommand{\ms}{\Delta m^2_{21}}
\begin{document}

\begin{flushright}
HRI-P-08-04-003\\
SISSA 27/2008/EP
\end{flushright}

\begin{center}
{\Large \bf Neutrino Oscillation Parameters After High Statistics 
KamLAND Results}
\vspace{.5in}

{ Abhijit Bandyopadhyay$^a$,
Sandhya Choubey$^{a}$,
Srubabati Goswami$^{a}$,
S.T. Petcov$^{b,c}$~
\footnote{Also at: INRNE,
Bulgarian Academy of Sciences, Sofia, Bulgaria},
D.P. Roy$^{d,e}$}
\vskip .5cm

$^{a)}${\small {\it Harish-Chandra Research Institute, Chhatnag Road, 
Jhunsi,
Allahabad  211 019, India}},\\
$^{b)}${\small {\it Scuola Internazionale Superiore di Studi Avanzati,
I-34014,
Trieste, Italy}},\\
$^{c)}${\small {\it INFN, Sezione di Trieste, Trieste, Italy}},\\
$^{d)}$
{\small {\it 
AHEP Group, Instituto de Fisica Corpuscular (IFIC), CSIC-U. de Valencia,
Edificio de  \\ 
Instituto de Paterna, Apartado de Correos 22085, E-46071 Valencia, Spain \\ 
$^{e)}$ Homi Bhabha Centre for Science Education, 
Tata Institute of Fundamental Research, \\Mumbai   400088, India}}

\vskip 1in

\end{center}
\begin{abstract}

We do a re-analysis to asses the impact of the 
results of the  Borexino experiment and the recent 2.8 KTy 
KamLAND data
on the solar neutrino oscillation parameters.  
The current Borexino results are found to have no impact on the 
allowed solar neutrino parameter space. 
The new KamLAND  data causes a significant reduction of the 
allowed  range of $\Delta m^2_{21}$, determining it with 
an unprecedented  precision of 
8.3\% at 3$\sigma$.
The precision of $\Delta m^2_{21}$ is  controlled practically  
by the KamLAND  data alone.  
Inclusion of new KamLAND  results also improves the upper bound on 
$\sin^2\theta_{12}$, but the precision of this parameter 
continues to be controlled 
by the solar data. The third mixing angle is constrained 
to be $\sin^2\theta_{13} < 0.063$ at $3\sigma$ from a combined 
fit to the solar, KamLAND, atmospheric and CHOOZ results. 
We also address the issue of how much further reduction of allowed range of 
$\Delta m^2_{21}$ and $\sin^2\theta_{12}$ 
is possible with increased statistics from KamLAND. 
We find that there is a sharp reduction of the  $3\sigma$ ``spread'' 
with enhanced statistics till about 10 KTy after which the 
spread tends to flatten out reaching to less than 4\% with 15 KTy data. 
For $\sin^2\theta_{12}$ however, the spread is more than 25\% even after 
20 KTy exposure and assuming $\theta_{12} < \pi/4$, as dictated by the
solar data.  We show that with a  
KamLAND like reactor ``SPMIN'' 
experiment at a distance of $\sim$ 60 km,    
the spread of $\sin^2\theta_{12}$ could be reduced 
to about 5\% at $3\sigma$ level while 
$\Delta m_{21}^2$ could be determined to within 
4\%, with just 3 KTy exposure.

\end{abstract}

\newpage

\section{Introduction}

Over the past few years there has been a 
paradigm shift in the studies of neutrino physics. 
The aim of neutrino experiments shifted from establishing 
the existence of neutrino mass and mixing to  precision determination 
of these oscillation parameters. In the 
case of solar neutrino oscillation, this
has been possible thanks to a succession of precision data from the SNO
and KamLAND experiments over the past few years. First, the simultaneous
measurement of solar neutrino events from both charged and neutral current
interactions by the SNO experiment 
\cite{Ahmad:2001an,Ahmad:2002jz}
 was instrumental in narrowing down the
solar neutrino mass and mixing parameters to the region 
of the so called Large Mixing Angle (LMA) solution 
\cite{snocc,snonc,snoccothers,snoncothers}.  
This was confirmed by the KamLAND reactor 
(anti)neutrino experiment \cite{kl162}. Moreover, it pinned 
down the solar neutrino mass parameter to two narrow 
bands called low-LMA and high-LMA (also called LMA-I and II, respectively), 
corresponding to 
the 1st and 2nd oscillation nodes \cite{kl162,solfit1,solfit2}. Then came 
the data from the second phase (salt phase) of SNO, 
which had a better detection efficiency for the 
neutral current events \cite{Ahmed:2003kj}. Including this data 
in a global analysis constrained the range of 
the solar neutrino mixing angle further, ruling out
maximal mixing at more than 6$\sigma$ 
level \cite{snosaltus,snosaltothers}. 
Besides, it strongly favoured the low-LMA
region of solar neutrino mass over the high-LMA, 
allowing the latter only at the
3$\sigma$ level. This was followed by the 766 Ty KamLAND data \cite{kl766}, 
which had a
nearly 5 times higher statistics than their first data. 
Including this data set in a
global analysis pinned down the 
solar neutrino mass finally to the low-LMA region,
while ruling out high-LMA at more than 4$\sigma$ level 
\cite{kl766us,sgnu04,kl766others}. 
In particular, our 
two-flavour neutrino oscillation analysis determined the 
best-fit solar
neutrino mass and mixing parameters to be $\dm = 8\times 10^{-5}eV^2$ and
$\sss = 0.28$, with a 3$\sigma$ spread of 
about 15\% and 30\% respectively \cite{kl766us}.
Extending this analysis to the three-flavour 
neutrino oscillation we found these
mass and mixing angle values to be robust. 
Finally, the three-flavour oscillation 
analysis led to a moderate improvement 
of the CHOOZ \cite{chooz} limit on the third mixing
angle, $\sin^2\theta_{13}$. It should be noted here that  
the most precisely determined neutrino parameter to date 
is the above mass parameter $\Delta m^2_{21}$; 
and the results from the KamLAND reactor 
neutrino experiment has played a pivotal role  in this.  

Recently the KamLAND experiment has published 
their 2.8 KTy data \cite{kl2008}, which 
increases the statistics of their 
earlier data by almost 4 times. Besides, they have 
reduced their systematic error and expanded 
the analysis to include the visible
energy range below 2.6 MeV. In this work 
we have updated our global 
analysis \cite{kl766us,sgnu04,sglp05,scnanp} with the inclusion of 
this new KamLAND data. As we shall see, its 
most important effect is a further reduction of the 3$\sigma$ spread
of $\Delta m^2_{21}$ by a factor of 2. We have also studied 
the effect of the first Borexino data \cite{borexino} 
on the result of this global analysis.

Section 2 is devoted to a two-flavour neutrino 
oscillation analysis of the global
solar neutrino data along with the new 
KamLAND reactor neutrino data. In Section 3
we extend this to a three-flavour neutrino oscillation analysis to check the
robustness of the oscillation parameters and also to update the limit on the
third mixing angle. 
In section 4 we study the impact of future data from Borexino
and KamLAND experiments on the precision of the solar neutrino mass and mixing
angle. We also discuss how the precision of this mixing angle measurement can
be improved dramatically by running a KamLAND type reactor 
SPMIN neutrino experiment
at a lower baseline length of 60 km \cite{th12}. 
We conclude by summarizing our main
results in section 5. 

\section{Two Flavour Neutrino Oscillation Analysis}

\begin{figure}[t]
\centerline{\psfig{figure=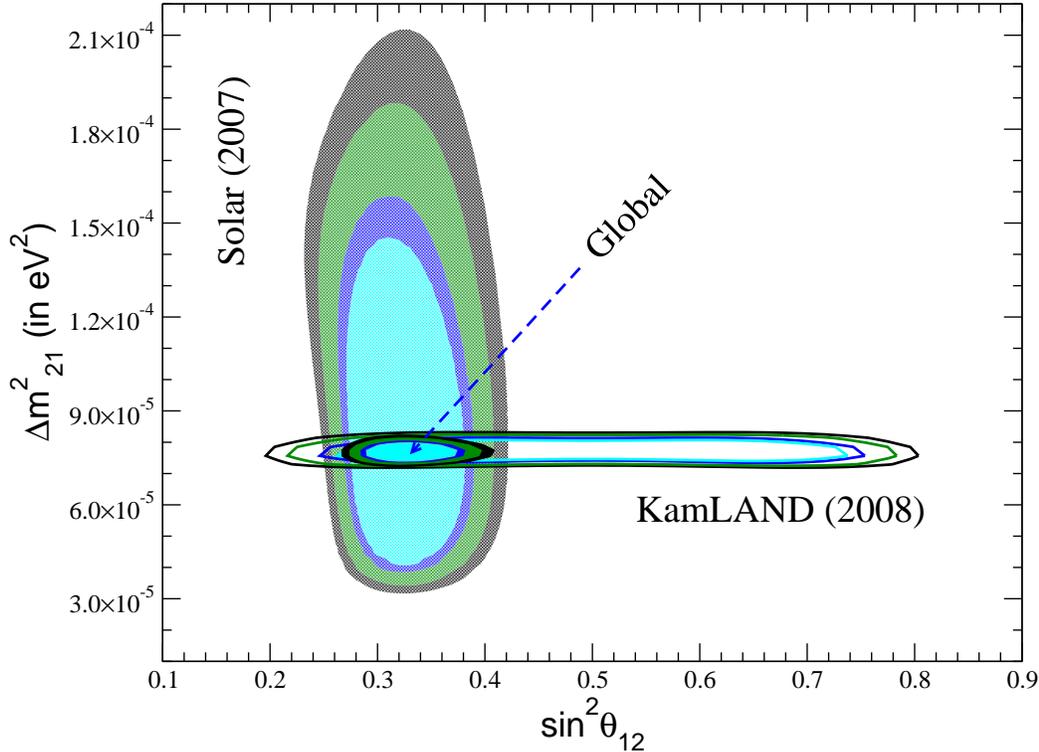,width=12.0cm,angle=270}}
\caption{\label{superpose}
The 90\%, 95\%, 99\% and 99.73\% C.L. 
allowed regions in the $\dm-\sss$ plane, 
obtained in a combined $\chi^2$-analysis of the global 
solar neutrino and the 2.8 KTy \kl spectrum data 
(shaded areas).
The regions allowed by the solar neutrino data and 
2.8 KTy KamLAND data  
are also shown separately.
} 
\end{figure}

\begin{figure}[t]
\centerline{\psfig{figure=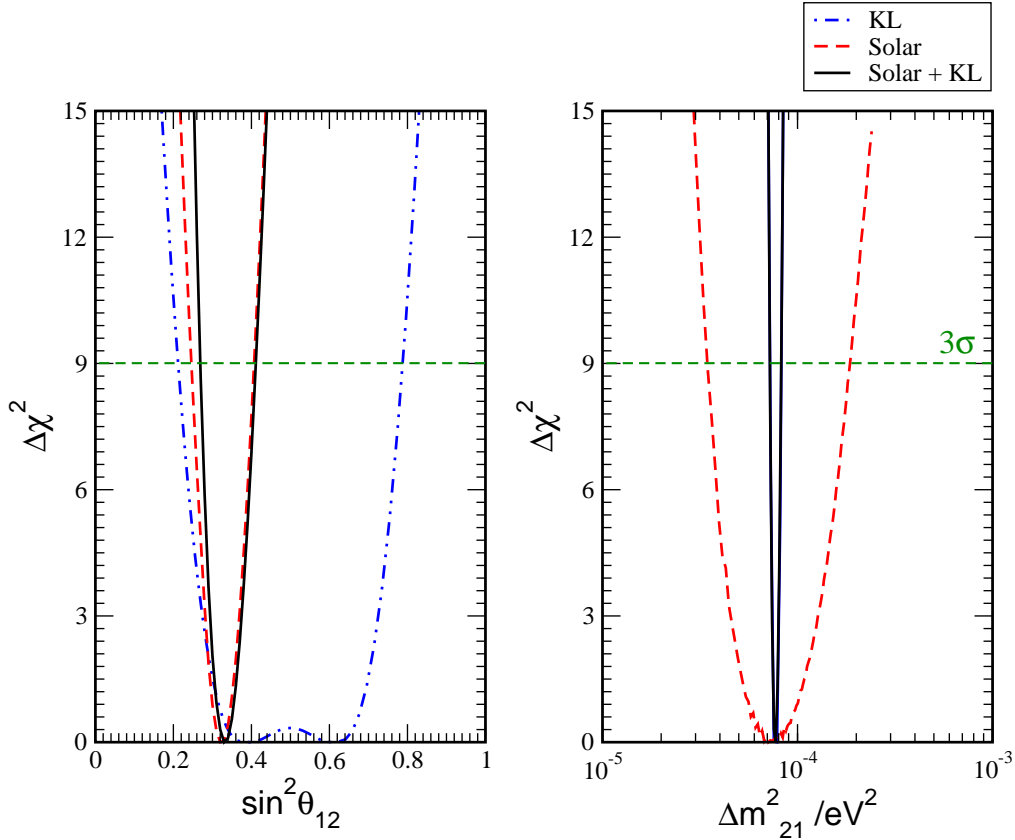,width=12.0cm,angle=270}}
\caption{\label{delvs12}
The $\Delta \chi^2 $ as a function of $\ms$ (right panel)
and $\sss$ (left panel). 
The results shown in both panels are obtained by allowing
all the other parameters to vary freely.
The dashed line shows the $3\sigma$ limit corresponding
to 1 parameter fit. The lines for only KL and solar+KL are 
indistinguishable in the right panel. 
}
\end{figure}

 We begin by reporting the status of the solar neutrino 
oscillation parameters $\ms$ and $\sss$. We present the 
allowed regions in the $\ms-\sss$ 
plane and 
investigate the impact of the new sets of results, {\it viz.}, 
the effect of adding the Borexino data, and 
the impact of the high statistics KamLAND results. 

\subsection{Oscillation Parameters from Solar Neutrino Data}

The first results from Borexino experiment were 
announced last year \cite{borexino} providing 
the first real time measurement of 
sub-MeV solar neutrinos. 
The observed rate is $47 \pm 7 (stat) \pm 12 (syst)$ /(day.100 ton) 
whereas the expected rate without oscillation is $75 \pm 4$/(day.100ton) 
according to the Standard Solar Model of  \cite{bp04}. 

This corresponds to an observed to expected Borexino 
rate of $R_B = 0.62 \pm 0.18$. We include this 
datum in our solar neutrino analysis and 
find the 90\%, 95\%, 99\% and 99.73\% allowed regions in the 
$\dm-\sss$ parameter space. These are shown as hatched contours  
in Fig. \ref{superpose}.  
We have used in this analysis 
the solar neutrino data on the total event rates 
from the radiochemical experiments, 
Chlorine (Homestake) \cite{cl} and Gallium 
(Gallex, SAGE and GNO combined) \cite{ga}, 
the 1496 day 44 bin Zenith 
angle spectrum data from SuperKamiokande \cite{sk},
and data from phase I  (pure $D_2O$ phase) \cite{Ahmad:2002jz} 
and phase II (salt phase) \cite{Ahmed:2003kj,Aharmim:2005gt} of the 
SNO experiment. For the SNO data set, we include 
the combined Charged Current (CC), 
Neutral Current (NC) and Electron Scattering (ES)
34 bin energy spectrum data from phase I  
and the 34 bin CC energy spectrum data (17 day bins and 17 night bins),
day and night NC rate data and day and night ES rate data 
from the
phase II. 
The $^8B$ flux normalization factor 
$f_B$ is left to vary freely in the analysis. 
For the other solar neutrino fluxes
($pp$, $pep$, $^7Be$, $CNO$, $hep$),
the predictions and estimated uncertainties 
from standard solar model (SSM) 
\cite{bp04} (BP04) have been utilized.
For further details of our solar neutrino 
code and error analysis we refer the reader 
to our earlier papers \cite{snocc,snonc,snosaltus}.

 We find that the present Borexino results make 
no impact on the allowed regions in the solar 
neutrino oscillation parameter space. 
The best-fit point from global solar neutrino 
data analysis stays unchanged at \cite{sglp05,scnanp}    
\begin{equation}
\Delta m^2_{21} = 6.4 \times 10^{-5}~{\rm eV^2},~~~
\sin^2\theta_{12} =0.33,~~~f_B=~0.84.
\label{eq2}
\end{equation}
%
These values of $\Delta m^2_{21}$
and $\sin^2\theta_{12}$
imply adiabatic MSW \cite{msw} conversions of 
the higher energy $^8B$ neutrinos
contributing to the SNO CC 
and SK event rates. The corresponding 
solar $\nu_e$ survival probability  is  given by 
$P_{ee}\simeq \sss$. 
For the low energy $pp$ neutrinos, 
which give the dominant contribution
to the signal in the Ga-Ge experiments 
(SAGE, GALLEX/GNO), the predicted 
$\nu_e$ survival probability 
is $P_{ee} =  1-0.5\sin^22\theta_{12}$. 
Using the indicated 
expressions for  $P_{ee}$, one can 
roughly check that the best-fit parameters given 
in Eq. (\ref{eq2}) provide  
an excellent 
fit to the global solar neutrino data. 
From an exact numerical analysis 
we obtain with a  $\chi^2= 114$ 
for 119 degrees of freedom.

To quantify the constraint the global solar neutrino data 
imposes on the parameters $\ms$ and $\sss$ individually, 
we show the $\Delta \chi^2$ as a function of these parameters in 
the right and left panels of Fig. \ref{delvs12}. 
Parameters which do not appear on the x-axis are left to 
vary freely in the fit. 
The red dashed lines correspond to 
the case where only solar neutrino results 
are included. 
The constraints on the 
individual oscillation parameters at any given C.L. 
for a one parameter fit can be read off from this figure. 
We give in the first row of 
Table \ref{spread} the ranges corresponding to 
the $3\sigma$ C.L. We also tabulate the corresponding 
``spread'' which quantifies the uncertainty on the 
given oscillation parameter and is defined as
\be
{\rm spread} = \frac{ prm_{max} - prm_{min}}
{prm_{max} + prm_{min}}\times 100,
\label{error}
\ee
%
where ${prm}$ denotes the parameter \dm or $\sin^2\theta_{12}$,
and $prm_{max}$ and $prm_{min}$ are the maximal and minimal values of
the chosen parameter allowed at a given C.L.
Solar neutrino results restrict $\sss$ to be 
uncertain at 3$\sigma$ by only $\pm 30\%$ 
around the best-fit, while for $\ms$ 
the 3$\sigma$ uncertainty is still as large as $\pm 70\%$. 
\begin{table}
\begin{center}
\begin{tabular}{ccccc}
\hline
{Data set} & (3$\sigma$)Range of & (3$\sigma$)spread in  & 
(3$\sigma$) Range of
&(3$\sigma$) spread in \cr
{used} & $\Delta m^2_{21}$ eV$^2$
& {$\Delta m^2_{21}$} & $\sin^2\theta_{12}$
& {$\sin^2\theta_{12}$} \cr
\hline
{only sol} & 3.0 - 17.0
&{70\%} & $0.21-0.39$ &30\%\cr
{sol+162 Ty KL}& 4.9 - 10.7 
& 37\%
& $ 0.21-0.39$ & 30\%  \cr
{sol+ 766.3 Ty KL}& 7.2 - 9.5
& 14\% &
$0.21-0.37$ & 27\% \cr
{sol+2.8 KTy KL} & $7.1 - 8.3$ & 7.8\% & 0.26 - 0.42 & 23.5\% \cr
{\it only KL} & ${\it 7.2 - 8.5}$ & \it{8.3\%} & \it{0.2 - 0.5} & \it{43\%} 
\cr
\hline
\end{tabular}
\caption{\label{spread}
3$\sigma$ allowed ranges of $\Delta m^2_{21}$ and
$\sin^2\theta_{12}$
from the analysis of the global solar neutrino, 
and global solar neutrino + \kl (past and present) data.
We show also the \% spread in the 
allowed values of the two neutrino oscillation 
parameters. Note that for only \kl we ignore the allowed 
region of $\sss$ in the Dark Zone ($\theta_{12} > \pi/4$) so that the 
maximum allowed value of $\sss$ is 0.5}
\end{center}
\end{table}

\subsection{Neutrino Oscillation Parameters from KamLAND Data Alone}

In their most recent paper, the KamLAND 
collaboration has made public,  
data corresponding to a statistics 
2.8 KTy \cite{kl2008}. 
The earlier data releases were for 0.162 KTy 
\cite{kl162} and 0.7663 KTy \cite{kl766}. 
Apart from an increased exposure time, 
the new data set is based on enlarged fiducial volume, 
full volume calibration to reduce the 
systematic error and expansion of 
the analysis to include the 
visible energy
\footnote{The visible 
energy is defined as $E_{vis}\simeq E_\nu - 0.8$ (MeV), where 
$E_\nu$ is the energy of the antineutrino.} 
spectrum below 2.6 MeV.  All these have been 
very important improvements, especially the 
measurement of the spectrum below 
2.6 MeV. The earlier two data sets from 
KamLAND were only for visible energy above 2.6 MeV, 
while the latest data set covers the entire available reactor 
spectrum, with threshold visible energy of 0.9 MeV. 
We use the 13 bin
\kl spectrum data with a threshold from 0.9 MeV and 
define a $\chi^2$ 
assuming a Gaussian distribution  as  
\be
\chi^2_{\mathrm KL} = \sum_{i,j=1}^N (R_i^{\rm expt}-R_i^{\rm theory})
(\sigma_{ij}^2)^{-1}(R_j^{\rm expt}-R_j^{\rm theory})
\label{chi2}
\ee
%
where $R^{\rm theory}$ and $R_i^{\rm expt}$ are the 
theoretically predicted and experimentally observed 
number of events in the $i^{th}$ energy bin, and 
$\sigma_{ij}^2$ is the error correlation matrix comprising of the 
statistical and systematic errors. The latter is taken to be 4.1\%, 
fully correlated between the energy bins.    
The other details of our analysis can be found in 
\cite{solfit1,kl766us,prekl}.
Some of the reactors, particularly 
the Kashiwazaki-Kariwa and Fukushima I and II reactor complexes, were 
partially/totally shut-down during some of the period of data taking 
in KamLAND. We have approximately 
taken into account this change in the flux due to the 
reactor shut-down using the plots showing the time variations 
of the number of fissions in a given reactor and 
hence the expected reactor $\anue$ flux in KamLAND \cite{talknove}.
We have also used the information on the reactor operation schedules 
available on the web \cite{web}. 

The 90\%, 95\%, 99\% and 99.73\% C.L. 
allowed areas 
in the $\ms-\sss$ parameter space, obtained 
using only the \kl data, 
can be seen within the
open contours in Fig. \ref{superpose}. 
We show the allowed 
regions derived from the solar neutrino 
and \kl data taken individually 
in the same plot to allow for better comparison. 
The best-fit point 
for the \kl data alone,
according to our analysis, is at 
\begin{equation}
\Delta m^2_{21} = 7.7 \times 10^{-5}~{\rm eV^2},~~~ 
\sin^2\theta_{12} =0.39~.
\label{eq:bf}
\end{equation}
%
We note that both these best-fit values are larger 
than those obtained from the analysis of 
the solar neutrino data only.
Note also that while the \kl data constrains $\ms$ much better 
than the solar neutrino data, 
the constraint on the mixing parameter $\sss$ 
from the solar neutrino data is much stronger. 
The range of allowed values for $\ms$ and $\sss$ at a given 
C.L. derived using the \kl data alone 
can be seen from the blue dashed lines in Fig. \ref{delvs12}. 
The limits at $3\sigma$ and the corresponding spread 
are given in Table \ref{spread}. 
The latest KamLAND data alone excludes the 
high-LMA solution at more than 4$\sigma$. Note that 
the earlier 766 Ty \kl results disfavored high-LMA at 2.56$\sigma$ 
only (1 parameter fit). 

\subsection{Constraints from Combined Solar and \kl Data Analysis}

  For the combined analysis of solar and \kl data we define the 
global $\chi^2$ as 
\be 
\chi^2_{global} = \chi^2_{\odot} + \chi^2_{KL}~,
\ee
%
where $\chi^2_{KL}$ is the $\chi^2$ for the \kl 
analysis given in Eq. (\ref{chi2}), and 
$ \chi^2_{\odot} $ is the $\chi^2$ computed from the 
global analysis of the world solar 
neutrino data. We refer the reader to our earlier papers 
\cite{snocc,snonc,snosaltus} for the details 
concerning $ \chi^2_{\odot} $. The results are plotted 
as C.L. contours shown by the shaded zones in Fig. \ref{superpose}. 
We find that with the inclusion of the latest 
\kl spectrum data, the allowed range of $\Delta m^2_{21}$ 
is sharpened considerably and the 
solar neutrino data plays practically 
no role in constraining $\Delta m^2_{21}$. 
On the other hand, the solar neutrino 
data is instrumental in reducing the 
allowed range of values of $\sin^2\theta_{12}$. 
The best-fit for combined solar neutrino and \kl 
data analysis is at,
\begin{equation}
\Delta m^2_{21} = 7.7 \times 10^{-5}~{\rm eV^2},~~ \sin^2\theta_{12}=0.33,~~
f_B = 0.84.
\label{eq3}
\end{equation}
%
The best-fit value of $\Delta m^2_{21}$ we find 
agrees very well with that obtained by the $\kl$ 
collaboration \cite{kl2008},
while our best fit value of $\sin^2\theta_{12}$ 
is somewhat lower than that found in \cite{kl2008} 
because of differences in the fitting procedure. 
The best-fit value of $\Delta m^2_{21}$ in the global fit is
controlled by the \kl data, 
whereas the best-fit value of $\sin^2\theta_{12}$ 
is controlled by the global solar neutrino data.  
For similar recent analyses see also \cite{kl2.8others}. 

The individual constraints on $\ms$ and $\sss$ 
from the combined analysis of the solar neutrino 
and \kl data can be seen in Fig. \ref{delvs12}, where we have 
plotted the $\chi^2 - \chi^2_{min}$ as a function of  
these parameters, taken one at a time. The corresponding 
$3\sigma$ allowed ranges and spread are given in 
Table \ref{spread}. In order to show how the 
statistics from the \kl experiment has effected the 
precision of the measurement of $\ms$ and $\sss$, 
we have also given in the Table the $3\sigma$ allowed ranges 
and spread we had obtained by combining the solar neutrino 
data with the first \kl results (0.162 KTy data) and 
second \kl results (0.7663 KTy data). We can see that 
while the error on $\ms$ has 
been dramatically reduced as 
\kl has accumulated more and 
more statistics, the uncertainty on $\sss$ has remained 
rather large. The reason why \kl has limited 
ability in constraining $\sss$ while its sensitivity to 
$\ms$ is quite remarkable
was pointed out in \cite{th12} 
and discussed in detail in \cite{shika,skgd,th12new,mina}. 

\section{Three Neutrino Oscillation Analysis}

\begin{figure}[t]
\centerline{\psfig{figure=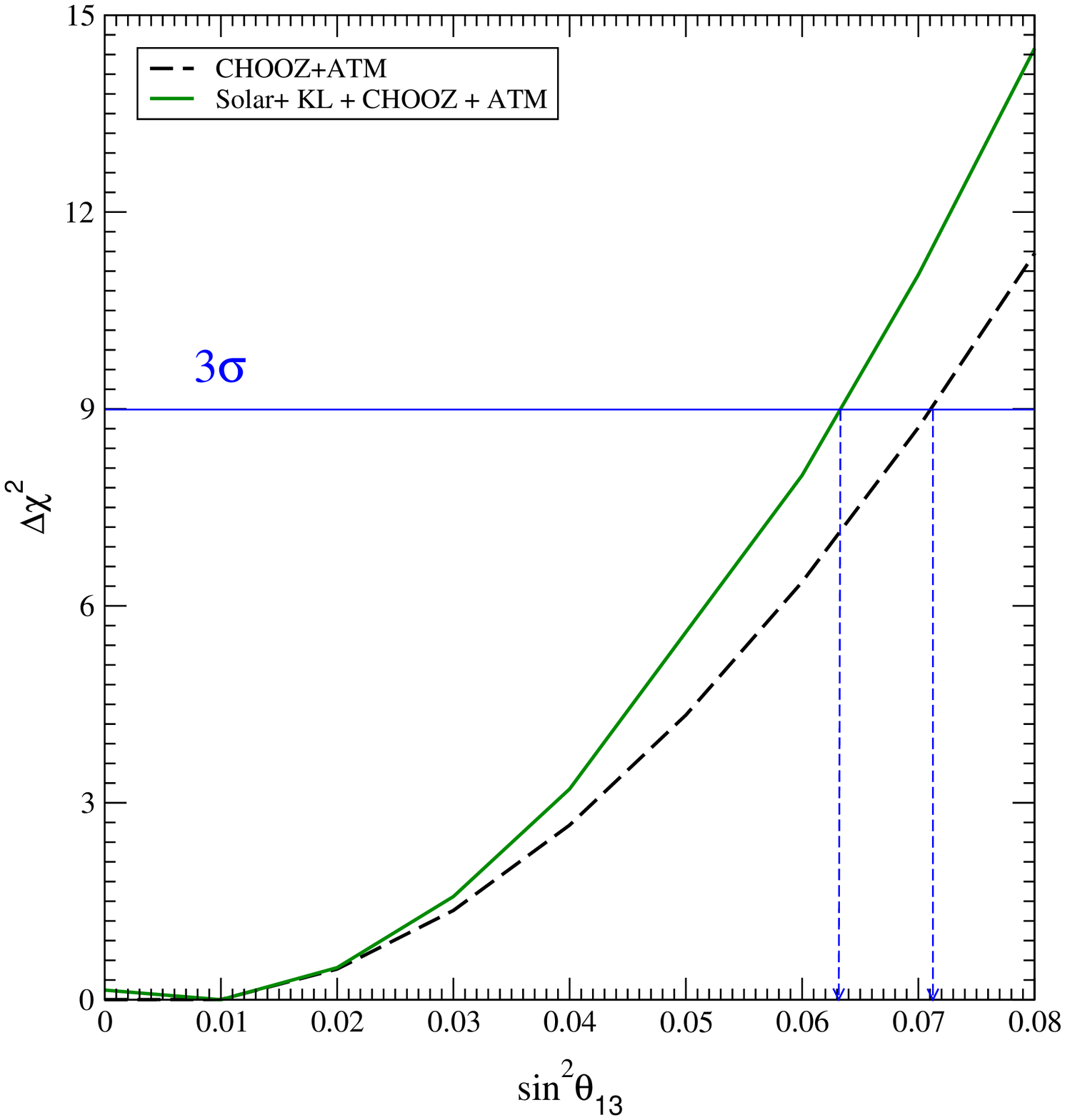,width=12.0cm}}
\caption{\label{delchith13}
Bounds on the mixing angle $\theta_{13}$ using 
the CHOOZ data only (dashed line)
and the combined solar, CHOOZ and \kl 
data (solid line). The $\Delta m_{31}^2$ is allowed to vary freely 
in its current $3\sigma$ limit allowed by the atmospheric 
and long baseline neutrino data. 
The short-dashed verticle lines show the $3\sigma$ limits 
corresponding to the case of
1 parameter fit. 
}
\end{figure}
%

\begin{figure}[t]
\centerline{\psfig{figure=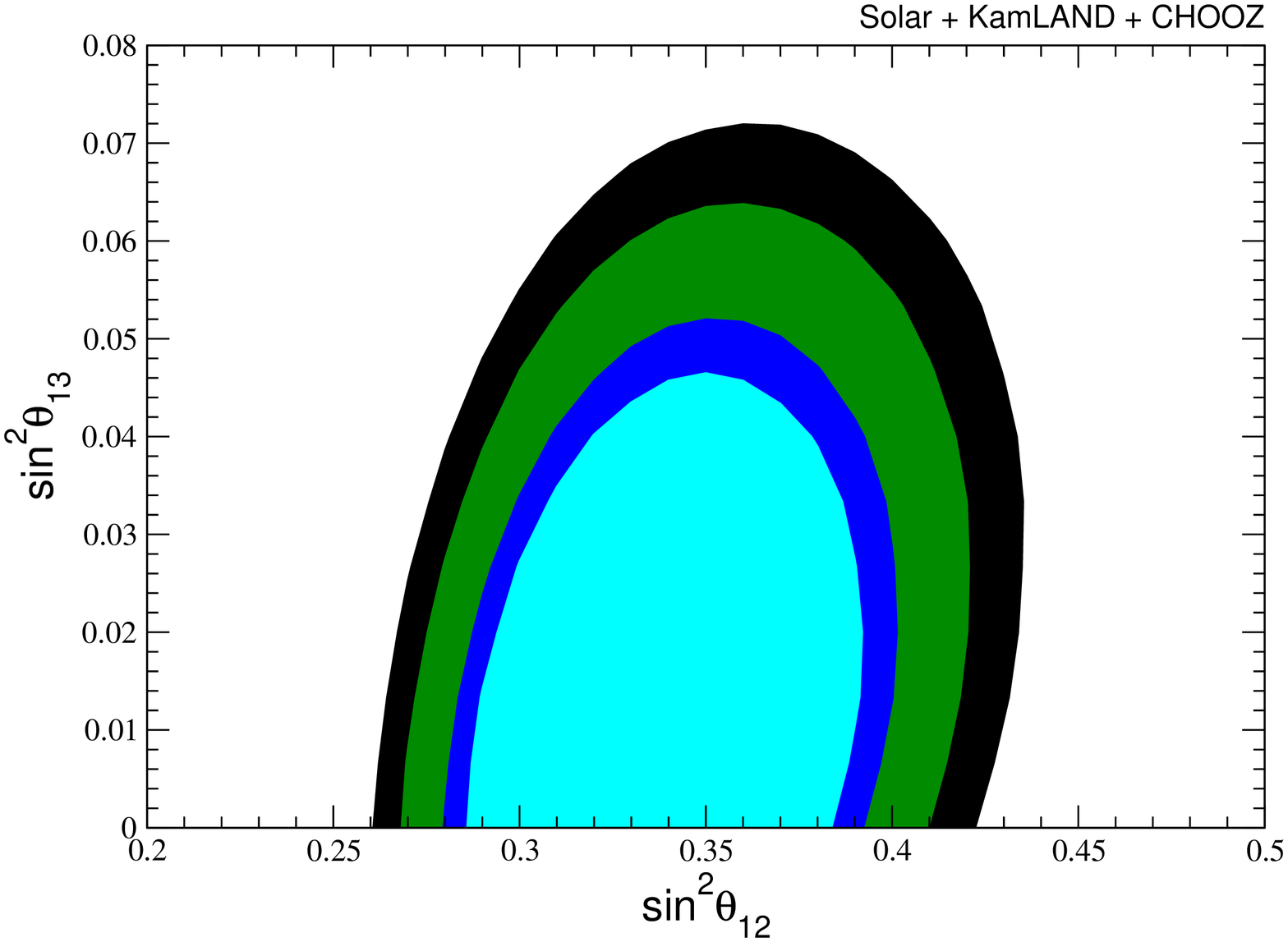,width=12.cm,angle=270}}
\caption{\label{solkl3osc}
The 90\%, 95\%, 99\% and 99.73\% C.L. allowed 
regions in the $\sss-\sin^2\theta_{13}$ plane, obtained 
in a three-neutrino oscillation analysis of the 
global solar and reactor neutrino data, 
including the data from the 
\kl and CHOOZ experiments.  
Here we use two parameter 
$\Delta \chi^2$ values to plot the C.L. contours.
}
\end{figure}

  So far we have restricted ourselves to two-generation oscillations 
where we have put the third mixing $\theta_{13}=0$. However, 
oscillation of solar and \kl (anti)neutrinos  
do depend on $\theta_{13}$, albeit weakly. 
Since $\Delta m_{31}^2 \gg \ms$,   
the three-neutrino oscillation
survival probability 
relevant for both solar and \kl (anti)neutrinos 
is  approximately given by
\be
P_{ee}^{3g} \simeq \cos^4\theta_{13}\,P_{ee}^{2g} + \sin^4\theta_{13}~,
\label{eq:3gen}
\ee
%
where $P_{ee}^{2g}$ is 
$\nu_e$ survival probability in the case
of two-neutrino 
oscillations. 
For solar neutrinos,
$P_{ee}^{2g}$ is given by the standard
expression (see \cite{SP88}), 
in which the electron number 
density $N_e$ is replaced by 
\cite{SP3nuosc88} $N_e\cos^2\theta_{13}$.
For  KamLAND, $P_{ee}^{2g}$ coincides with the 
usual two-neutrino vacuum oscillation probability used in the 
previous section. Thus, both solar and \kl have some 
sensitivity to $\theta_{13}$ and can therefore constrain 
it. We show in Fig. \ref{delchith13} the $\chi^2$ 
obtained as a function of $\sin^2\theta_{13}$ when all other 
oscillation parameters are allowed to vary freely. While 
$\ms$ and $\sss$ are allowed to take any value in fit, the values  
of $\Delta m_{31}^2$ are restricted within 
its current $3\sigma$ range.
We show 
results for analysis of the CHOOZ reactor antineutrino 
and atmospheric results  (solid line), 
as well as by adding solar and  \kl data  to this set (dashed line).
The combined  global data from solar neutrino, atmospheric neutrino  
and reactor antineutrino experiments put a bound 
of $\sin^2\theta_{13} < 0.063$ at $3\sigma$. 
We have checked that 
there is practically no increase
in the allowed regions in the $\ms-\sss$ plane,  
when one goes from two to three flavor
neutrino oscillation analysis of the global solar neutrino
and \kl spectrum data. 
To show the impact of the solar and \kl data on three neutrino 
parameters we present 
in Fig. \ref{solkl3osc} the 90\%, 95\%, 99\% and 
99.73\% C.L. allowed contours in the 
$\sin^2\theta_{12}-\sin^2\theta_{13}$ plane obtained from the 
combined analysis of the global solar neutrino data, the latest 
\kl data and CHOOZ data. 
It is to be noted that 
the $P_{ee}^{2g}$ for high energy $^8{B}$ neutrinos is $\sim f_B \sss$ 
while for \kl it is given as  
$1 - \sin^2 2\theta_{12} \sin^2\ms L/4E$ . 
Thus while for solar neutrinos an increase in $\theta_{13}$ 
implies an increase in $\theta_{12}$, for \kl an increase in $\theta_{13}$ 
would imply a decrease 
in $\theta_{12}$ \cite{sgsmirnov}.
This opposing trend 
is instrumental in putting 
constraints in the $\sin^2\theta_{12} - \sin^2\theta_{13}$ plane. 

\section{What lies in the Future }

\begin{figure}[t]
\centerline{\psfig{figure=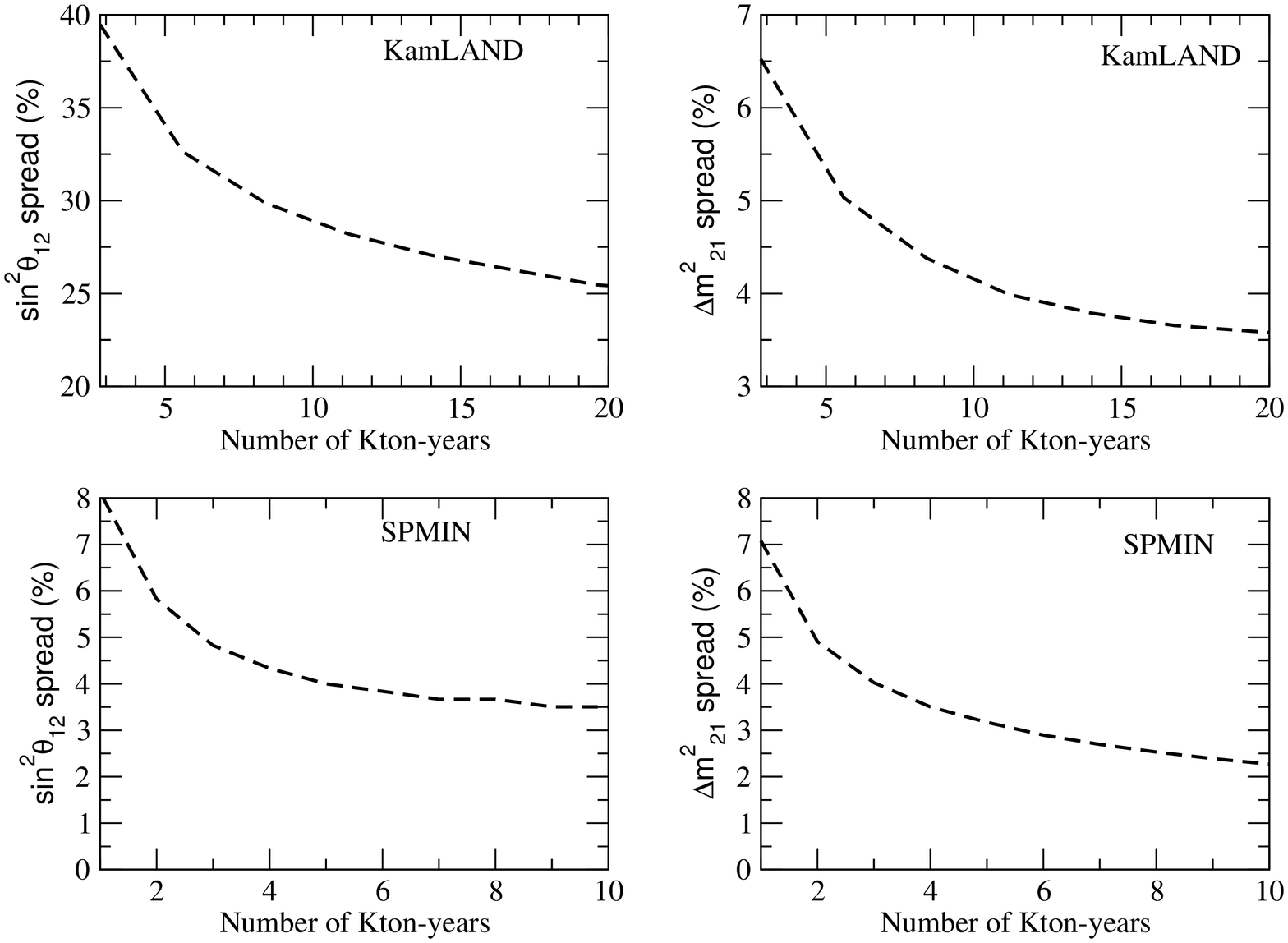,width=12.0cm,angle=270}}
\caption{\label{statistics}
Expected $3\sigma$ 
spread of $\ms$ and $\sss$ as a function of the statistics 
for KamLAND (upper panels) and the SPMIN experiment (lower panels). 
}
\end{figure}
%
  The field of solar neutrino research 
has become quite mature now. 
The latest results from Borexino experiment 
has made real time detection of 
the $^7{Be}$ solar neutrinos possible and 
the results are consistent with the 
expectations from the LMA solution. 
The results from the \kl reactor data have provided independent 
and solid support to the LMA solution of 
the solar neutrino problem. With the recent \kl 
data,  the precision of $\Delta m^2_{21}$ 
gets controlled solely by KamLAND.  
At this point we ask the question, what 
will be the impact of future results 
from Borexino  and KamLAND. 
In particular, we address two questions: 
\begin{itemize} 
\item Can improved precision of Borexino data 
play any role in further 
reducing the allowed ranges of 
$\ms$ and/or $\theta_{12}$? 
\item What will be the impact of a further 
increase of statistics of the \kl data? 
\end{itemize}  

To address the first point we analyze the solar neutrino data 
taking the Borexino rate as its present experimental value, 
but reducing the
1 sigma error (combined statistical and systematic) from 30 to 15\%. 
However, even then there is no impact of 
Borexino on the allowed solar 
neutrino parameter space. 
To asses the impact of the central value of the Borexino rate 
on the above result,  
we vary the allowed parameters in the combined solar 
and \kl analysis within their 3$\sigma$ range and use the maximum 
and minimum predictions for the Borexino rate as the central value 
and accomplish an analysis of the combined solar data using 
15\% total error. But 
the allowed parameter space in 
the $\Delta m^2_{21} -\sin^2\theta_{12}$ plane 
remains stable against these variations. 
However, the measurement of the 
$^7$Be neutrino flux with a higher precision 
will be very important 
for the determination of some of the 
basic solar model parameters \cite{s17}.

In order to address  the second question, we show in upper 
panels of Fig. 
\ref{statistics} the spread in $\sss$ (left panel) and 
$\ms$ (right panel) as a function of the number of KTy of data 
in KamLAND. The x-axis starts from the current \kl statistics 
of 2.8 KTy.   Note that while plotting the 
spread of $\sss$,  we ignore the allowed range of 
$\sss$ in the dark zone ($\theta_{12} > \pi/4$), as 
dictated by the solar data. 
The figure shows that the spread in $\ms$ shows a steady  
decrease till about 10 KTy of 
statistics of \kl after which the 
spread starts to decrease more gradually 
reaching to less than 4\% with 
15 KTy of statistics. 
The figure reveals that the spread in $\sin^2\theta_{12}$ from \kl 
also reduces with statistics, but even with 
20 KTy of data,  
the spread in $\sss$ is more than 
25\%, which is not significantly 
better than the value of 30\%  obtained from the current solar 
data (cf. Table 1). 
It has been already pointed out in the literature that 
maximum precision in $\sss$ can be obtained in a
reactor antineutrino experiment, identical to \kl in all
respects, except that the baseline of this experiment would be
tuned to the Survival Probability MINimum (SPMIN) \cite{th12,th12new,mina}.
Note that the present \kl experiment is situated at an 
average distance of about 180 km, which is a  
maxima of the survival probability (SPMAX). 
In the lower panels of this figure we show the 
projected sensitivity to these parameters in a ``SPMIN'' 
experiment \cite{th12,th12new,mina}. 
For the 
current best-fit $\ms$, the baseline  corresponding to 
SPMIN  would be at about $L=60$ km. 
One can see from the figure the remarkable sensitivity that this 
experiment would have to the mixing angle $\sss$. Even with 
1 KTy of data, we could determine $\sss$ to $\pm 8\%$ 
precision and this could improve to about 5\% with about 3 KTy 
of statistics. The sensitivity to $\ms$ is also seen to be good.
Although the survival probability is larger at the 
SPMAX than at the SPMIN,  
the latter is situated at a shorter distance of  60 km as 
compared to SPMAX (180 km at the present best fit value).  
So the distance factor makes up for the probability. 
Also it is to be noted that since \kl receives 
flux from several 
reactors at different distances, it 
is actually at an average SPMAX and 
so it cannot see the full distortion of the spectral shape. 
For the above  reasons a dedicated SPMIN 
experiment also gives a comparatively better sensitivity 
to $\Delta m^2_{21}$. We could determine $\ms$ within 
$\pm 4$\% precision with 3 KTy data. 
The above results are obtained by taking $\sin^2\theta_{13}=0$. 
However, inclusion of a non-zero 
$\sin^2\theta_{13}$ is not expected to 
alter the conclusions significantly \cite{th12new}. 
Another experimental idea which could be used to return 
very good precision to the solar neutrino oscillation 
parameters consists of doping the SuperKamiokande with 
gadolinium \cite{skgd,gadzooks}. 
 
\section{Conclusions} 

We have updated the solar neutrino parameter space including  the 
Borexino results and the 2.8 KTy \kl spectrum data 
in global solar neutrino 
oscillation analysis. 

The present 
Borexino results are found to have no impact on the solar neutrino 
parameter space. We also find that the allowed area 
in $\ms-\sss$ plane remains stable against reduction in Borexino error 
by half its present value or by shifting the central value within 
the predicted 3$\sigma$ range of the global solar and \kl analysis.  
The inclusion of the latest \kl results on the other hand 
causes a reduction in the spread in $\ms$ by a factor of 2.  

The allowed range of $\Delta m^2_{21}$ is   
controlled practically by the \kl data. 
There is also a slight increase in the lower bound of $\theta_{12}$ 
with the inclusion of \kl data,  though the precision in $\sss$ 
is controlled by the solar data. 

The $3\sigma$ upper limit on $\sin^2\theta_{13}$ 
from global solar, atmospheric and reactor antineutrino data is 0.063. 
There is practically no change 
in the allowed region in the $\ms-\sss$ plane 
when one goes from two to three flavor
neutrino oscillation analysis of the global solar neutrino 
and \kl spectrum data. 
The effect of combined solar and reactor antineutrino data on three flavour 
parameters have been presented in terms of allowed regions in
the $\sss -\sin^2\theta_{13}$ plane.

We also studied the impact of further reduction of \kl statistics 
on the precision of $\ms$ and $\sss$ and find that till about 10 KTy of 
statistics there 
is steady improvement of precision beyond which the spread in $\ms$  
flattens out, reaching less than 4\% with 15 KTy of statistics. 
Spread in 
$\sss$ shows hardly much improvement 
with increased \kl statistics. 
Even after accumulation of 20 KTy of statistics, the spread hovers 
around  
25\%, which is not much better than 
the 30\% precision which the current solar 
data gives. A dramatic improvement in precision in $\sss$ is possible in 
a dedicated \kl type of experiment at a distance of 60 km. Such an experiment 
can give 5\% precision in $\sss$ and 4\% precision in $\ms$  
with only 3 KTy of statistics.

\vskip 8pt
The work of A.B., S.C. and S.G. was supported by the Neutrino Project 
under the XIth plan at Harish Chandra Research Institute. 
D.P.R was supported in part by BRNS(DAE)
through Raja Ramanna Fellowship and 
in part by MEC grants FPA2005-01269, SAB2005-0131.
The work of S.T.P. was supported in part by  
the Italian INFN and MIUR programs 
``Fisica Astroparticellare'' and
 ``Fundamental Constituents of the Universe'',
and by the European Network of Theoretical 
Astroparticle Physics ILIAS/N6 
(contract RII3-CT-2004-506222).



\end{document}